# Designing Hierarchical Exploratory Experiences for Ethnic Costumes: A Cultural Gene-Based Perspective

Ma Xiaofan[†]; Yan Lirong[†]; Zhao Weijia[†]; Zeng Weiping[†]; Wu Huiyue\*

School of Journalism and Communication, Sun Yat-sen University, Guangzhou, China
\* wuhuiyue@mail.sysu.edu.cn
[†] These authors contributed equally to this work

Ethnic clothing is a vital carrier of cultural identity, yet its digital preservation often results in static displays that fail to convey deep cultural meaning or foster user engagement. Existing practices lack a systematic design framework for translating the hierarchical cultural connotations of these garments into dynamic, personalized, and identity-promoting digital experiences. To address this gap, this paper proposes a novel Three-Layer Cultural Gene Framework that systematically decodes ethnic costumes from their surface-level visual symbols, through their mid-level socio-cultural contexts, to their inner-layer spiritual core. Based on this framework, we designed and implemented an interactive digital platform featuring two key innovations: a "gene-first" exploratory path that encourages curiosity-driven discovery, and an AI-powered co-creation experience. This generative feature allows users to co-create personalized narratives and images based on their understanding of the "inner-layer" genes, transforming them from passive observers into active co-creators. A mixed-methods user study (N=24) was conducted to evaluate the platform. The findings demonstrate that our approach effectively enhances users' cultural cognition, deepens their affective connection, and significantly promotes their sense of cultural identity. This research contributes a validated framework and a practical exemplar for designing generative, identity-building digital experiences for cultural heritage, offering a new pathway for its preservation and revitalization in the digital age.

## 1 Introduction

Propelled by globalization and technological advancements, the digital preservation of cultural heritage, such as ethnic costumes, has become a significant topic of global concern. The concept of "cultural genes" is understood as the basic unit determining the inheritance and variation of cultural heritage; possessing traits of heredity and mutability, they are not only a vital mechanism for cultural continuity but also the foundation for cultural innovation (Wang, 2003). In this context, ethnic costumes serve as significant carriers of a nation's cultural genes. They are laden with rich historical information and cultural connotations, and they perform the vital function of embodying ethnic identity and facilitating cultural transmission (Hu, 2001).

The development of digital technology has created new opportunities for the innovative design and dissemination of cultural genes. The digital translation of these genes has become an effective pathway for cultural preservation and innovation. Through modern technical means, the cultural genes of ethnic costumes can be translated and presented to better showcase their underlying historical and cultural value, thereby promoting their innovative design and widespread dissemination (Luo et al., 2023). However, digitalization has introduced a "paradox of preservation": on one hand, it solves the challenge of preserving and displaying traditional costumes, which are often made of fragile materials, through digital archiving; on the other hand, many digital practices are limited to high-fidelity replication or static information displays, creating a "glass showcase" effect in the digital world. This model struggles to inspire deep user participation and fails to convey the rich cultural connotations and affective value behind the garments (Sun et al., 2024).

A limitation of existing research lies in the absence of a systematic design framework to guide the effective translation of the hierarchical cultural connotations of ethnic costumes—from their surface-level symbols to their deep spiritual core—into a dynamic, personalized digital experience that can foster identity. To address this challenge, this study proposes and develops an interactive digital platform centred on the theory of "cultural genes." We argue that to achieve meaningful cultural transmission, it is necessary to move beyond mere visual display and, instead, to design the relationship between users and the connotations of cultural heritage—a relationship built on "exploration, experience, and co-creation."

To this end, we first constructed a "Three-Layer Cultural Gene Analysis Framework" to systematically decode ethnic costumes, from their surface-level visual symbols and mid-level socio-cultural contexts to their inner-layer spiritual core. Based on this framework, we designed and implemented a web-based digital platform. The platform's core innovation is twofold: firstly, it provides a "from-point-to-surface," non-linear exploration path, where users can start from a micro "cultural gene" (such as a pattern) to autonomously discover and construct knowledge; secondly, it introduces Generative Artificial Intelligence (AIGC) as a "co-creative partner in cultural narrative," allowing users, based on their understanding of the inner-layer genes, to co-create personalized stories and images with AI to establish deep affective connections. To validate the effectiveness of this design framework and platform, we conducted a mixed-methods user study with 24 participants.

The core contribution of this research lies in proposing an operational Three-Layer Cultural Gene Model for the digital translation of the cultural connotations of ethnic costumes. This theoretical framework directly guided our subsequent design practice: we developed and implemented a digital platform that integrates a "gene-first" exploration path with an AI co-creation experience, providing a new practical exemplar for revitalizing cultural heritage. The effectiveness of this platform—in terms of sparking exploratory interest, deepening cultural understanding, and promoting cultural identity—was preliminarily validated through a user study involving 24 participants.

The remainder of this paper is structured as follows: Section 2 reviews related work and the theoretical foundation. Sections 3 and 4 detail the design and implementation of our three-layer cultural gene framework and the digital platform. Sections 5 and 6 present the design and results of our user study. Section 7 discusses the implications, limitations, and future work. Finally, Section 8 concludes the paper.

## 2 Theoretical foundation

### 2.1 Cultural Genes: A Theoretical Lens for Analyzing Heritage

The concept of the "Cultural Gene," or "Meme," originates from biology and describes the basic unit of cultural information that replicates and evolves in a manner similar to biological genes (Dawkins, 1976). Given China's profound and continuous cultural heritage system, this concept offers a particularly suitable theoretical lens for systematically analyzing the internal mechanisms of its cultural transmission. In applying this theory to the local context, Chinese scholars have further defined the cultural gene as the core factor determining the inheritance and variation of a cultural system—an underlying principle or spirit inherent in cultural phenomena, capable of being transmitted across time and space (Wang, 2003; Bi, 2001).

In the field of design research, the analysis of cultural elements often draws on semiotic perspectives, dividing them into different layers. This approach coincides with the classic "explicit" and "implicit" dichotomy in cultural gene research (Zhao, 2008). Explicit Genes, analogous to the "denotation" of a symbol (Barthes, 1967), constitute the directly perceptible form of a cultural product. Presented through visual elements like patterns and colors, they are the primary objects of design translation. Implicit Genes, on the other hand, are closer to the "connotation" of a symbol, involving deep cultural values, social norms, and belief systems. They cannot be directly observed and must be interpreted and understood through a "Thick Description" of the cultural context (Geertz, 1973). This hierarchical structure, from explicit to implicit, provides the foundation for systematically decoding the cultural genes of ethnic costumes.

In design practice, the core method for applying cultural genes is the Translation of Cultural Genes. This definition transcends simple replication, pointing instead to a dynamic design process: creatively reinterpreting and re-encoding a culture's core elements, features, or spiritual values into forms adapted for new media, technologies, and contexts (Zhao, 2021). This study focuses on the cultural genes within ethnic costumes, which, as information patterns formed in specific regional and cultural environments, are inheritable, identifiable, and a key manifestation of ethnic style (Chu, 2015). As carriers of these cultural genes, ethnic costumes are not only the external material manifestation of their cultural connotations but, more importantly, a key medium for constructing and expressing Ethnic Identity (Jiang et al., 2023). Therefore, to achieve a meaningful digital translation of these genes, establishing a hierarchical analytical framework capable of systematically decoding and organizing them is crucial.

In existing research, researchers have indeed made attempts to build such frameworks. Despite varied perspectives, a commonality emerges when researchers translate the cultural genes of traditional ethnic costumes: they often employ a three-layer framework, such as "matter-form-spirit" or "surface-deep-ultra-deep symbols" (Song & Zhan, 2020; Wang & Bai, 2019). Although these translation frameworks may target different ethnic groups or costume types, they all reflect a logical progression from the surface to the core and from the concrete to the abstract, thus providing multi-dimensional theoretical support for the translation of cultural genes in ethnic costumes.

While these frameworks provide valuable theoretical support for the translation of cultural genes in ethnic costumes, our review reveals two critical research gaps, particularly at the intersection of cultural theory and HCI design practice.

Firstly, existing frameworks are largely confined to the static deconstruction of surface-level cultural symbols, primarily for the purpose of guiding the design of physical products. There has been significantly less discussion on how to translate these hierarchical genes into dynamic and interactive digital experiences. Consequently, many digitalization projects for ethnic costumes are limited to high-fidelity 3D replications or static information displays, failing to effectively convey the cultural connotations of the middle and inner layers.

Furthermore, current research often focuses on the analysis of a single ethnic group or region, lacking a systematic and cross-cultural approach to translation. This fragmented perspective makes it difficult to distill common principles, hindering the development of a more holistic theoretical framework for cultural identity. This limitation is particularly salient in the digital age, where the potential of emerging technologies like Generative AI and Virtual Reality to create innovative and culturally-aware experiences remains underexplored. Persisting with this isolated and static paradigm risks diminishing the modern vitality of cultural heritage, as it lacks the dynamic interaction mechanisms necessary to foster deep and broad cultural identity.

Therefore, there is an urgent need to construct a systematic framework that integrates the dynamic characteristics of multi-ethnic cultural genes with digital translation pathways. Such a framework would lay the foundation for new mechanisms of cultural exploration and experience, ultimately promoting the formation and deepening of cultural identity.

## 2.2  The Formation of Cultural Identity: From Exploration to Commitment

Cultural Identity refers to the affirmation and sense of belonging an individual or group feels towards a shared culture, serving as the foundation for social integration, interaction, and consensus (Hall, 1996). The complex and dynamic process of identity formation involves the continuous interplay of three dimensions: cultural exploration, practical experience, and cognitive commitment (Phinney & Ong, 2007).

Marcia's (1966) identity status theory first established "commitment" as a core dimension of identity development, defining it as an individual's adherence to a chosen identity. In the context of cultural identity, commitment manifests as an individual's strong sense of identification and belonging to a particular culture, a willingness to internalize it as their own cultural identity, and to enact its norms and values in daily life (Yan & Dong, 2023). Building on Marcia's foundation, Crocetti et al. (2008) further proposed a three-factor identity model, which clearly defines and refines the concepts of "Broad Exploration" and "In-depth Exploration." Broad Exploration refers to the extent to which individuals search for and engage with diverse cultural elements and values *before* making a cultural commitment. In-depth Exploration, conversely, emphasizes the process of continuously and deeply reflecting on and evaluating the historical context, intrinsic meanings, and practical aspects of one's *existing* commitment (Crocetti et al., 2008; Crocetti, 2017).

It's noteworthy that broad and in-depth exploration are not isolated processes; their synergy jointly promotes a deeper understanding and affirmation of cultural identity. Broad exploration provides individuals with diverse cultural information, helping them form initial impressions through contact and preliminary cognition. In-depth exploration, through further cognitive processing, critical reflection, and emotional investment, prompts individuals to deeply understand and consolidate their existing commitments (Meeus, 2011). In this process, commitment is both a motive for and an outcome of exploration, serving as an internalized goal (Crocetti et al., 2014). This continuous cycle of exploration, reflection, and commitment constitutes the key psychological mechanism of cultural identity construction (Umaña-Taylor et al., 2014).

However, the exploration of cultural information alone is insufficient to foster a profound cultural identity; it requires a **meaningful cultural experience** as a key catalyst. Existing research indicates that a cultural experience involves not only the cognitive absorption of content but also emphasizes behavioural participation and emotional resonance within authentic or simulated contexts (Falk & Dierking, 2016). In the fields of HCI and cultural heritage, a meaningful experience transcends simple information Browse. As Sun et al. (2024) propose in their IVR study, such experiences must be constructed through embodied motivation and affective connectedness. Similarly, researchers also emphasized promoting cultural participation by transforming users from passive observers into active creators (Wang et al., 2023). During this process of cultural experience, individuals, through affective connection and the reconstruction of meaning, can more actively internalize cultural values, thereby reinforcing their commitment and further deepening their cultural identity (Bruner, 1990; Falk & Dierking, 2016).

Therefore, cultural exploration and cultural experience do not have a simple linear relationship; rather, they form a synergistic and interdependent system that jointly contributes to the complex mechanism of identity formation. While a theoretical basis for identity formation exists, how to effectively integrate the processes of cultural exploration and experience into the digital interactive environment for appreciating ethnic costumes—in a way that dynamically promotes the generation and deepening of users' commitment to their culture—remains an area requiring further in-depth investigation.

### 2.3 The Digitalization of Ethnic Dress: Current Practices and Core Challenges

Currently, the exhibition and dissemination of cultural heritage are undergoing a profound digital transformation and innovation (Kalay, 2007). This trend has also impacted the field of ethnic costumes, where the research perspective has gradually shifted from a singular focus on material preservation to a deeper exploration of cultural connotations and innovation in digital experiences (Jiang et al., 2023; Zhao & Wang, 2011; Yu & Zhu, 2024). Museums, as core venues for cultural exploration and identity formation, have played a pivotal role in this transformation. Especially in the post-pandemic era, cultural institutions are actively exploring new paradigms for display and communication (Mo et al., 2023). An increasing number of museums are leveraging digital technologies to pioneer new trends in exhibition design, including multi-sensory interaction, immersive experiences, and non-linear narratives, to enhance the interactivity and affective resonance between exhibits and users (Wang et al., 2023).

In the domain of ethnic costumes, institutions such as the Palace Museum, the China National Silk Museum, and the Beijing Institute of Fashion Technology have set precedents by creating new models for digital costume appreciation through digital exhibition halls and online platforms (Wei et al., 2019). With the development of technologies like AR, VR, and MR, a greater variety of immersive experiences have emerged. A notable example is VisHanfu (Yu et al., 2024), an interactive VR system designed to promote Hanfu knowledge. By focusing on the unique cross-shaped flat structure of Hanfu, the system allows users to freely observe and manipulate digitally restored models through various interactions, including "Rotate," "Drag," "Unfold," and "Try-on."

Despite the significant achievements of current digital practices for ethnic costumes in preservation and display, critical issues emerge when examined from the perspective of exploration and experience. Many digital platforms remain in a mode of static replication and information display. Users are often limited to passively Browse high-resolution images or 3D models online, an experience akin to looking through a glass showcase in the digital world, which lacks meaningful interaction. This static digital exhibition model has been shown to be relatively inefficient in stimulating public interest and understanding (Wang et al., 2023).

Furthermore, even in many excellent and highly interactive experiences, the underlying design often relies on pre-defined, linear narratives, which can limit user participation. Even in cases of non-linear storytelling, user choices are frequently confined to content branches pre-determined by curators (Wang et al., 2023). The user's role is more that of an "explorer" than a "creator." This one-to-many broadcast-style of information delivery struggles to adapt to the diverse knowledge backgrounds, interests, and emotional needs of individual users.

Many technological applications remain at the level of "technological spectacle," failing to achieve deep integration with cultural connotations and lacking a systematic approach to the entire experience process (Wang et al., 2023). This results in experiences that may be novel, but are often fragmented and shallow in their effectiveness at fostering the internalization of cultural knowledge and the establishment of affective connections (Sun et al., 2024). These limitations not only reflect the shortcomings of current practices but also highlight the urgent need for innovative methods of presentation.

The rise of Generative Artificial Intelligence (AIGC) presents a groundbreaking opportunity to overcome these challenges. It enables a potential paradigm shift from "developer-scripted narratives" to a new model of "user-AI co-created stories." AI technology can lower the barrier to content creation, giving general audiences the opportunity to become creators of cultural content, thereby enhancing their sense of participation and cultural belonging (Wang, 2023). While this technology is currently applied mainly in the field of fashion design within the ethnic costume domain (Mendoza et al., 2023), its prospects in cultural dissemination have garnered widespread academic attention. The highly customizable nature of generative technology opens up new possibilities for the digital display of ethnic costumes, representing a technological breakthrough that provides new research directions for cultural inheritance and innovation.

Therefore, the core research question of this study is: **How can we innovate exploration and experience methods based on a hierarchical cultural gene model to promote users' identification with ethnic costume culture?**

This paper argues that by designing a digital platform that supports the hierarchical exploration of ethnic costume culture through the translation and presentation logic of cultural genes, and by integrating personalized experiences driven by generative AI, we can help users build a stronger sense of cultural identity. The specific contributions of this paper are as follows:

1. Drawing on existing cultural gene research and the specific characteristics of ethnic costumes, we propose a hierarchical cultural gene translation framework that encompasses a surface, middle, and inner layer.
2. We developed and implemented a generative AI digital experience platform. This platform combines identity formation mechanisms with the multi-level translation and presentation of cultural genes to promote deep, interactive engagement with cultural identity.
3. We offer reflections on the future possibilities for museums and HCI researchers to integrate the concept of cultural genes into digital museum collections.

## 3  Methodology: The Three-Layer Cultural Gene Model

### 3.1  The Three-Layer Cultural Gene Model: Surface, Middle, and Inner Layers

Drawing upon our earlier discussion of explicit and implicit genes, this study constructs a Three-Layer Model for analyzing and translating the cultural genes of ethnic costumes. The model's structure is inspired by established hierarchical approaches in the field of cultural product analysis. For example, numerous scholars have discussed the layered nature of design translation (Xu et al., 2023), and Leong and Clark (2003) previously classified the connotations of cultural products into three levels: a tangible, material outer layer; a behavioral, customary middle layer; and an ideological, intangible inner layer. Inspired by this, and tailored to the unique characteristics of ethnic costumes, we propose a more focused analytical framework consisting of a Surface Layer, Middle Layer, and Inner Layer (see Figure 1). This framework aims to systematically decode a garment, progressing from its most intuitive visual form, through its socio-functional context, to its deeply embedded spiritual core.

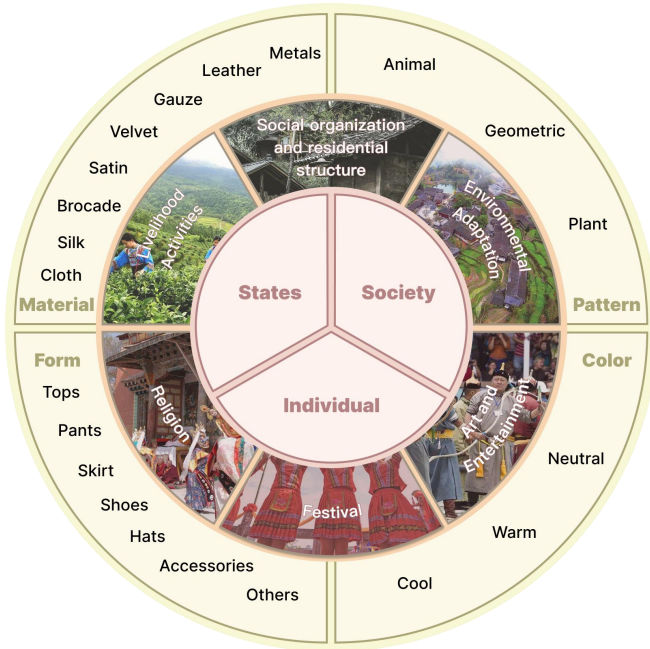

*Figure 1: The Three-Layer Model of Cultural Genes in Ethnic Costumes (from outer to inner: Surface Layer, Middle Layer, and Inner Layer).*

### 3.1.1 Surface Layer

The Surface Layer consists of the most intuitive and easily identifiable Explicit Genes. It is primarily presented through four visual and physical characteristics: Pattern, Color, Material, and Form. This layer forms the basis of a user's first impression of a costume and is the primary object of digital archiving and design re-creation. As demonstrated in numerous digital museum practices, the systematic cataloging and metadata standardization of these surface elements is the cornerstone for building searchable and researchable digital collections (Goodrum & Martin, 2005).

In this study, we meticulously classified these four features based on authoritative sources:

- **Pattern:** We adopted the established classification system from an authoritative dictionary of Chinese fine arts, dividing patterns into three main categories: geometric, animal, and plant.
- **Color:** Based on color theory, these were classified as cool, warm, and neutral.
- **Material:** Our classification covers nine typical materials, including cloth, silk, brocade, satin, velvet, gauze, leather, and metal.
- **Form:** We referenced the classification methods of a leading university museum in China renowned for its ethnic costume collection, categorizing items by structure and function, such as tops, pants, skirts, shoes, hats, and accessories.

### 3.1.2 Middle Layer

The Middle Layer serves as the Cultural Context Layer, connecting the surface symbols to the inner spirit. It aims to answer "why" a garment is designed in a certain way and "how" it is used in social life. This layer is implicit, as it cannot be fully understood by merely observing the garment itself. As revealed by Jiang et al. (2023), the innovation of handicrafts is intrinsically linked to their social environment. Drawing upon foundational works in the field, such as Research on Chinese Ethnic Minority Costume Culture, this study summarizes the middle-layer genes—which involve behavioral perceptions and lifestyles—into six core dimensions: Religious Beliefs, Festive Ceremonies, Social Structures, Livelihood Activities, Arts & Entertainment, and Environmental Adaptation. On our digital platform, this layer of genes is translated and presented through narrative text, historical images, and archival footage.

### 3.1.3 Inner Layer

The Inner Layer is the Spiritual Core of the cultural genes in ethnic costumes. It represents the implicit, collective Worldview, Community Ethics, and Life Aspirations that a group has formed over its long history. These deep genes are often unspoken but are profoundly encoded in the materials, colors, patterns, and structure of the garments through myths, legends, beliefs, and rituals. Decoding this layer aims to touch upon the most fundamental values of the culture. In this study, we systematically organize these rich spiritual connotations into three guiding dimensions: Values at the State Level, Guiding Principles at the Societal Level, and Moral Norms at the Individual Level. The specific cultural concepts and their interpretations within each dimension are detailed in Table 1. These thematic dimensions are intertwined, collectively forming the rich and profound spiritual world of ethnic costumes. On our digital platform, this layer of genes is primarily translated and conveyed through user-participated, AI-assisted generative narrative experiences.

| Level | Core Cultural Concept | Expression Examples in Ethnic Costume | Cultural Connotation & Contemporary Relevance |
|---|---|---|---|
| **Values at the State Level** | Prosperity | The recurring use of patterns symbolizing fertility and abundance (e.g., pomegranates, fish, wheat ears), and the use of splendid materials in festive and wedding attire. | Materializes the collective aspiration for life's continuity, abundant resources, and well-being. This universal pursuit inspires innovation in contemporary cultural industries. |
| | Democracy | Certain ceremonial garments worn by elders or community leaders may symbolize a tradition of collective deliberation and respect for members' rights. | Reflects the inherent wisdom in community governance and social harmony. This provides insights into diverse forms of social organization and inspires modern approaches to participatory consensus-building. |
| | Civility | Exquisite craftsmanship, complex narrative patterns, and strict dress codes for specific rituals all demonstrate a high regard for wisdom, skill, and behavioral propriety. | Highlights the universal human respect for knowledge, skill, and decorum. This inspires a greater appreciation for the transmission of traditional crafts and mutual respect in cross-cultural communication. |
| | Harmony | The extensive use of natural materials, colors, and motifs (flora and fauna) embodies the philosophy of "harmony between heaven and humanity" and a reverence for nature. | Emphasizes the valuable ecological wisdom inherent in many traditional cultures. In an era of global environmental challenges, this concept offers profound inspiration for sustainable design. |
| **Guiding Principles at the Societal Level** | Freedom | The structure of nomadic attire, often designed for ease of movement, reflects an adaptation to a migratory lifestyle and a yearning for physical and spiritual freedom. | Reflects the human desire for vitality, mobility, and liberation of the spirit. This spirit encourages the exploration of the unknown and continues to inspire artistic and cultural expression. |
| | Equality | Symbolic elements shared across different genders or social classes in ceremonial attire; patterns narrating stories of resistance against oppression. | Expresses the pursuit of fairness and the inherent value of individuals within the community. These ideas contribute to the development of more inclusive societies today. |
| | Justice | The solemn and symmetrical design of garments worn by law-keepers or ritual hosts may symbolize the authority and responsibility to uphold community norms and dispense fairness. | Embodies the universal human need to establish and maintain a just order. Studying traditional norms helps us understand the formation of justice concepts in diverse cultural contexts. |
| | Rule of Law | The specific rules governing who wears what on which occasion is itself a form of social contract, reflecting a shared respect for communal order and established customs. | Emphasizes the importance of abiding by agreements and respecting rules in social life. This consciousness is a cornerstone of social stability and cultural transmission. |
| **Moral Norms at the Individual Level** | Patriotism / Community Guardianship | Totemic patterns symbolizing a specific region or ethnic group; iconic garments that evoke collective emotion and are worn during significant community events. | Reflects a deep emotional bond and sense of responsibility towards one's homeland and community. This identity is the foundation of cultural diversity and a spiritual tie that unifies a community. |
| | Dedication | The meticulous, time-consuming, and highly skilled craftsmanship involved in making a garment is itself a testament to the artisan's dedication, patience, and creative spirit. | Showcases the universal virtue of creating value through diligent work and skill. This "spirit of craftsmanship" is a vital driving force for social and cultural development. |
| | Integrity | The simple, unadorned style of certain garments may symbolize an honest and upright character; the formality of attire worn for oaths or important agreements implies a high regard for trustworthiness. | As the foundation of interpersonal relationships and social trust, integrity is a crucial virtue shared across cultures, especially vital for building harmonious communities in our complex modern world. |
| | Friendliness | The vibrant colors and welcoming motifs (e.g., blooming flowers) of festive or ceremonial attire often convey signals of hospitality and inclusiveness. | Embodies the universal human desire to establish friendly relations and foster emotional communication. This quality is essential for promoting cross-cultural understanding and cooperation. |

*Table 1. A Three-Level Interpretation of the Inner-Layer Cultural Genes in Ethnic Costumes*

## 3.2 Data Acquisition and Multi-Level Annotation Based on the Three-Layer Model

The data acquisition for this study began with the screening of authoritative data sources. We established contact with a top-tier university museum in China, renowned for its massive collection of ethnic costumes. After formal communication and obtaining research authorization from the institution, we were granted access to its digital collection, which includes 852 costumes and a total of over 15,000 images and texts. The authority and richness of this data source provided a solid material foundation for our research.

To systematically transform this raw material into data suitable for analysis and design, we followed the three-layer cultural gene model previously constructed, conducting a rigorous multi-level data annotation and analysis process. First, we used the official text descriptions provided on the museum's website as the ground truth for annotating the surface-layer genes—such as pattern, material, and form—as well as the middle-layer cultural contexts. Two researchers independently "transcribed" the features of each garment into our classification system based on the extracted official text. The results were then exchanged for a double-coder cross-check; any inconsistent entries were resolved by a third researcher to ensure the accuracy and authority of the annotations.

For the extraction of the color gene, we employed a two-step process combining computational analysis with manual classification. We applied the K-Means clustering algorithm to perform a cluster analysis on the pixels of each costume image to extract the dominant hue (the cluster with the largest proportion), and recorded its objective Hexadecimal color code (Hex code). Subsequently, researchers then mapped these color codes to the perceptual categories of cool, warm, or neutral based on color theory. This rigorous process, which combines authoritative sources, cross-validation, and computational analysis, allowed us to convert unstructured image and text information into a structured gene database with rich metadata, laying a highly reliable data foundation for the subsequent design of the exploratory experience.

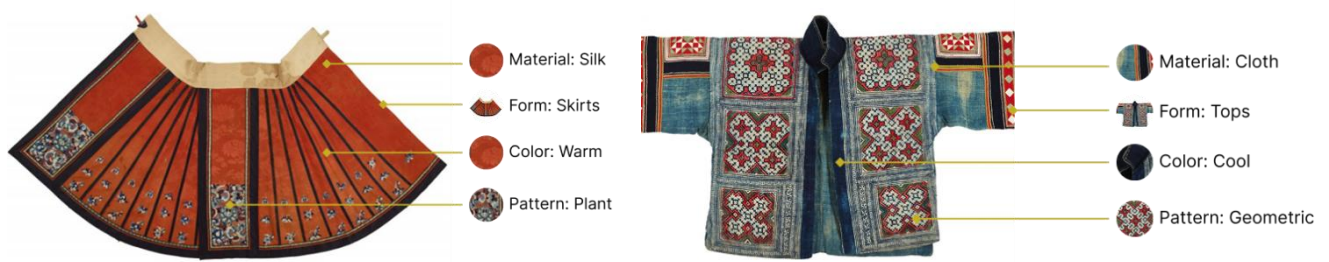

*Figure 2: Annotation of surface-layer genes on two examples of ethnic costumes: a brick-red jacquard silk skirt with "Sanlan" seed embroidery, and a Miao women's jacket from Guizhou with batik and appliqué.*

# 4 A System for Exploring Cultural Genes through Generative Experience

This chapter delves into how our digital system, through a hierarchical design based on cultural genes, promotes user exploration and enhances cultural identity. The core of the design revolves around a progressive cultural experience path, aiming to guide users from a surface-level to a deep-level understanding, ultimately achieving comprehensive cultural comprehension and affective resonance.

## 4.1 Design Rationale and Process

Our research aims to put the previously constructed Three-Layer Cultural Gene Model and the theories of identity formation into practice. Through innovative interaction design, we guide users to complete a cultural experience journey from the surface to the core—from exploration to identity—fostering an in-depth engagement with ethnic costumes. To achieve this goal, our design adheres to several core principles.

1. **Supporting Layered Exploration** - ensuring that the system's interaction maps to the surface-middle-inner structure of cultural genes to provide a clear path from board exploration to depth exploration. Users first explore surface genes, Browse tags and filtering to enter costume themes that pique their interest. They then proceed to detail pages for a deeper dive into the cultural context and connotations, culminating in the generation of narrative texts and contextual images.
2. **Enabling Co-creation of Meaning** - The system is designed to transcend simple information display by empowering users to participate in the personalized creation of cultural narratives through generative AI, transforming them from observers to co-creators.
3. **Providing Generous Interfaces** - Drawing on the concept by Whitelaw (2015), the system offers diverse starting points, allowing users to initiate their journey from a micro cultural gene (such as a pattern) based on personal interest. This is supported by a responsive layout for clear information delivery and dynamic interaction design to enhance the immersive experience.

Following these principles, we adopted an iterative design process. Over a 16-week design cycle, we conducted five rounds of optimization. This process began with competitive analysis and user interviews to define core functions and information architecture (Round 1), followed by the creation of low-fidelity wireframes for preliminary usability testing (Round 2). We then completed the high-fidelity UI visual design, referencing the styles of institutions like the Palace Museum and adding innovative elements such as carousels and transition animations to enhance aesthetic appeal and cultural relevance (Round 3). Subsequently, we developed an interactive prototype with a focus on refining the prompt engineering for the AI generation module (Round 4). The final round involved a comprehensive user study to gather feedback for final optimization (Round 5). Through this rigorous process, we finalized the technical implementation and design scheme, ensuring the platform could not only provide intuitive navigation but also delve deeply into the cultural value of ethnic costumes, thereby offering users a more comprehensive and meaningful exploratory experience.

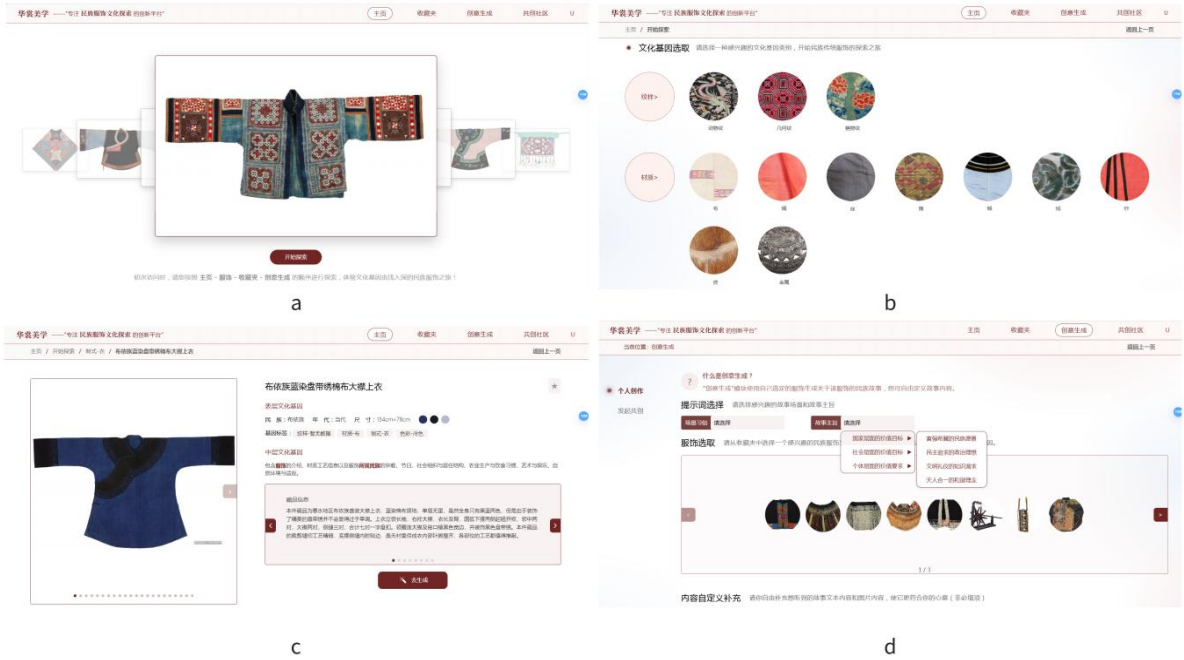

*Figure 3: Selected interfaces of the digital platform: (a) Homepage; (b) Directory of surface-layer gene tags; (c) Costume detail page; (d) Creative generation page.*

## 4.2 System Architecture and Core Features

Integrating the three-factor model of cultural exploration, our system constitutes a multi-faceted platform that brings together tag-based display, story generation, image creation, and multimodal presentation (see Figure 4). The system leverages a layered logic to integrate cultural genes, providing users with a progressive exploratory path from intuitive surface elements to the inner spiritual core. This design not only enhances the perceptual depth of ethnic costume culture but also stimulates affective identity through multi-dimensional interaction.

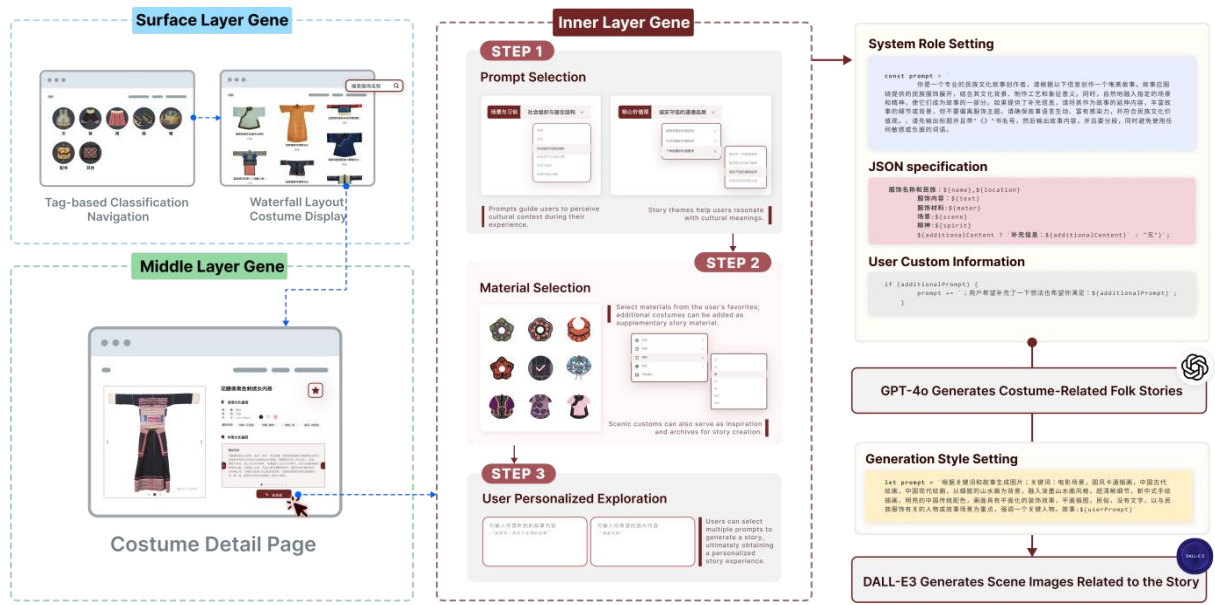

*Figure 4: The architecture of the exploratory experience for cultural genes in ethnic costumes.*

**Stage 1: Breadth Exploration of the Surface Layer**

As the starting point for cultural exploration, the core design of this stage is to break from the traditional viewing model that centers on complete garments. Instead of Browse an entire piece of clothing directly, we innovatively establish independent "cultural genes"—such as a single pattern, a type of material, a specific color, or a structural form—as the fundamental units and entry points for exploration.

This "from-point-to-surface" perspective aims to guide users to focus on details that are easily overlooked in a holistic viewing. For instance, a user might notice the subtle evolution of a specific pattern (e.g., a butterfly motif) across the costumes of different ethnic groups and eras, or the unique texture and luster of a special material (e.g., cloud-patterned brocade). We argue that this focus on detail is more effective at sparking user curiosity and providing richer, non-linear pathways for subsequent exploration.

To support this philosophy, the platform offers two complementary exploration modes to serve different types of Breadth Exploration behaviors:

- For users with no clear goal who prefer serendipitous discovery, the platform presents surface genes (patterns, colors, materials, forms) visually through tag-based classification and navigation. Users can browse openly, a "from-point-to-surface" method designed to spark serendipity and the joy of exploration.

- For users with a clear objective who wish to perform a targeted query, the platform provides a keyword search function. Users can input specific costume names, ethnic groups, or particular cultural genes (e.g., "dragon pattern," "silk") to quickly locate all relevant entries.

These two modes collectively create a space for breadth exploration that allows for both free-roaming and precise targeting. Regardless of the path taken, the user ultimately arrives at a costume's detail page to access high-definition images and basic information, thereby establishing a preliminary cognitive understanding.

**Stage 2: In-depth Exploration of the Middle Layer**

Building upon this foundation, the middle-layer cultural genes are presented through the display and interpretation of cultural details. The platform integrates multi-dimensional information—such as the costume's historical background, craft techniques, ritual functions, and social customs—into the detail page in a narrative format. This translation and presentation of cultural context are designed to immerse the user in the ethnic culture and their own reflections, further encouraging active participation in the generative experience and fostering a deeper sense of cultural identity. Users can also save items of interest to a "favorites" collection for revisiting, and these selected costumes can be directly used in the creative generation module.

**Stage 3: Generative Experience of the Inner Layer**

To make the abstract inner-layer genes tangible, the system employs Generative AI technology. We designed a Prompt Engineering system that incorporates costume information, middle-layer cultural context, and user selections. Users can choose story themes and scenic customs that resonate with them (focusing on the costume's function in religious, festive, or artistic contexts). The system combines these inputs with backend data, calls upon GPT-4 to generate a culturally rich and emotionally resonant ethnic story, and uses the DALL-E 3 model to create a matching visual image. Through this multimodal presentation of text and image, the user engages in a personalized, co-created dialogue with the inner-layer genes in a virtual context, thereby profoundly deepening their cultural identity and affective commitment.

In summary, this design transforms cultural genes into perceptible forms, providing users with a progressive exploratory path to deepen their cultural identity. Through this multimodal interaction, users can gradually appreciate the unique value of ethnic cultures, ultimately forming a positive sense of cultural identity and commitment.

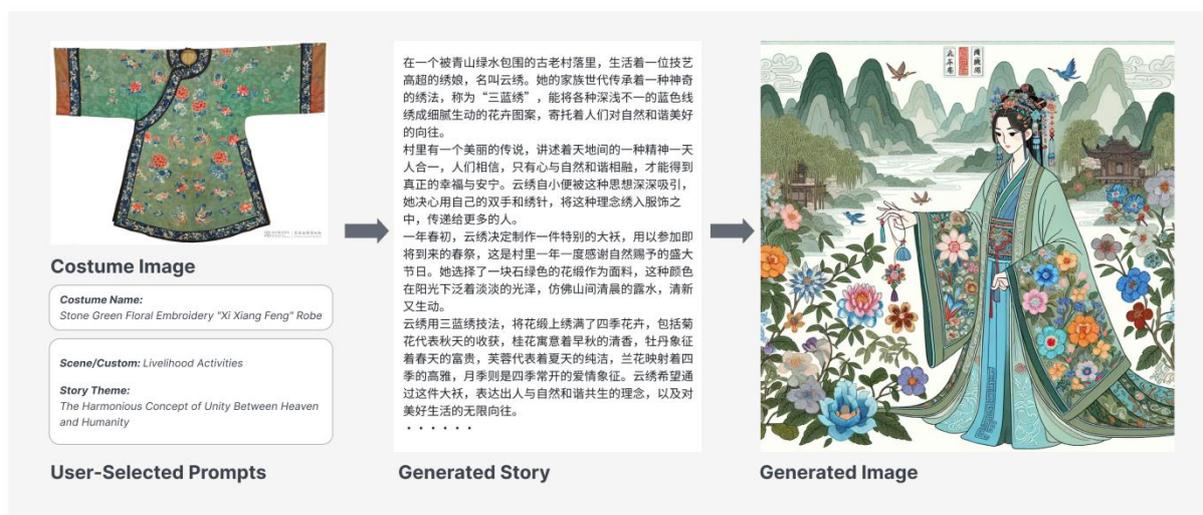

*Figure 5: The process of generating stories and images, showing the subject, user prompts, the generated story, and the resulting image.*

## 5 User Study

### 5.1 Study Design and Goals

The primary goal of this study was to evaluate whether the digital platform, through its hierarchical cultural gene translation framework and interactive mechanisms, could guide users to a progressively deeper understanding of the different layers of ethnic costume culture, evoke affective resonance, and ultimately enhance their sense of cultural identity. This objective was to be assessed through quantitative and qualitative data analysis, evaluating the platform's design effectiveness and its role in promoting cultural identity, thereby providing an evidence-based foundation for future system optimizations.

### 5.2 Participants

We recruited a total of 24 participants (11 female, 13 male) from a comprehensive university in China. The participants' ages ranged from 19 to 28 (M = 23.5, SD = 2.1). All held a bachelor's degree or higher and came from diverse academic backgrounds. Among them, three participants had an ethnic minority background.

In pre-study interviews, although the majority of participants (21) reported having encountered ethnic costumes through travel or museum visits, they generally acknowledged that their understanding was superficial, lacking in-depth knowledge of specific details and historical context. This finding highlights the potential value of our research in providing an experience for deep cultural exploration and education.

### 5.3   Apparatus and Materials

The study was conducted in a university's human-computer interaction laboratory. Participants accessed our developed web-based digital platform on a 13-inch MacBook Pro equipped with the latest version of the Chrome browser. The entire session was recorded using professional audio equipment to ensure the integrity and accuracy of the interview data.

The research materials primarily consisted of two parts:

- **Digital Platform Prototype:** A fully functional web application embodying the core designs of this research, including the Three-Layer Gene Model, the hierarchical experience exploration path, and the AI generation feature.

- **Questionnaire:** A comprehensive questionnaire combining scales from the Technology Acceptance Model (TAM) and the Multigroup Ethnic Identity Measure-Revised (MEIM-R) for quantitative evaluation.

### 5.4   Procedure

Each participant's session lasted approximately 40 minutes. First, a researcher gave a brief introduction to the platform's functions and the procedure, ensuring the participant understood the tasks. This was followed by a 20-minute free exploration period, allowing the participant to familiarize themselves with the platform's features and content. Next was a 10-minute specific task period, during which the researcher asked the participant to complete at least one "ethnic costume story and image generation" task. After completing all operations, participants spent about 5 minutes filling out an online questionnaire. Finally, the researcher conducted a 5-10 minute semi-structured interview based on the participant's questionnaire answers and their interaction with the platform.

### 5.5   Data Collection and Analysis

We collected data through two methods: questionnaires and interviews. For quantitative data, we primarily collected ratings based on a 7-point Likert scale. This data was used to analyze users' evaluations of the platform in terms of Perceived Ease of Use (PEoU), Perceived Usefulness (PU), and its impact on promoting cultural identity. For qualitative data, the interview recordings were transcribed verbatim, and we employed Thematic Analysis to code and identify recurring themes in participants' feedback regarding the platform's strengths, weaknesses, their user experience, and cultural engagement.

# 6 Results

## 6.1 Quantitative Analysis: Usability, Engagement, and Cultural Identity

We conducted a statistical analysis of the 24 completed questionnaires. The results indicate a positive overall reception of the digital platform we designed. The quantitative results for key dimensions are presented as follows:

| Dimension | Item | Mean (M) | Std. Dev. (SD) |
|---|---|---|---|
| **Perceived Ease of Use (PEoU)** | The classification and presentation are clear and easy to understand. | 6.13 | 1.03 |
| **Exploratory Usefulness** | Exploring via costume elements increased my interest and curiosity. | 6.13 | 0.9 |
| | The detail page provided comprehensive and rich information. | 6.08 | 0.72 |
| | The information helped me better understand the ethnic culture. | 6.04 | 0.81 |

*Table 2: Core Platform Experience Ratings (N=24)*

| Dimension | Sub-dimension | Item | Mean (M) | Std. Dev. (SD) |
|---|---|---|---|---|
| **Cultural Identity** | **Affective Commitment** | The experience deepened my sense of cultural identity. | 6.08 | 1.25 |
| | | The generated images deepened my emotional connection. | 5.13 | 1.48 |
| | | I could feel the spiritual value conveyed by the costumes. | 5.71 | 1.08 |
| | **Behavioral Exploration** | I am more willing to spend time learning more about ethnic cultures. | 6 | 1.14 |
| | | I am willing to recommend this website to friends and family. | 6 | 1.22 |
| | | I am willing to share the AI-generated content on social media. | 5.83 | 1.46 |

*Table 3: Cultural Identity Dimension Ratings (N=24)*

| Dimension | Item | Mean (M) | Std. Dev. (SD) |
|---|---|---|---|
| **Generative Feature Usefulness** | The generated content helped me better understand the cultural background. | 5.33 | 1.63 |
| | The generated content enhanced my interest. | 5.46 | 1.56 |
| | Interacting with AI content deepened my understanding of the costumes. | 5.33 | 1.71 |

*Table 4: Generative Feature Usefulness Ratings (N=24)*

The data shows that the platform received high ratings in terms of usability, its ability to spark exploratory interest, and its effectiveness in promoting cultural identity.

Most critically, the data shows a very positive signal in the dimension of Cultural Identity. High scores were given by participants in both the dimension of 'Affective Commitment,' which reflects internal emotions (e.g., the experience deepened their sense of identity, M=6.08, SD=1.25; the images enhanced their emotional connection, M=5.13, SD=1.48), and the dimension of 'Behavioral Exploration,' which reflects external behavioral tendencies (e.g., willingness to spend more time learning, M=6.00, SD=1.14; willingness to recommend to friends and family, M=6.00, SD=1.22).

Overall, the quantitative analysis results provide preliminary validation for the effectiveness of our design. The data indicates that our platform is not only a usable and engaging tool for exploration but, more importantly, that it holds the potential to positively impact users' sense of cultural identity through its unique, AI-integrated exploratory path. These findings will be further corroborated and elaborated upon in the following section on our qualitative analysis.

### 6.2 Qualitative Findings: Journey from Exploration to Affective Commitment

Through a Thematic Analysis of the semi-structured interviews, we found that the participants' experience journey clearly mapped onto our "from-surface-to-core" design. The findings reveal how the platform performed in sparking exploration, guiding understanding, facilitating co-creation, and ultimately, fostering cultural identity.

6.2.1  The "Gene-First" Approach as an Effective Catalyst for Breadth Exploration

Participants generally expressed appreciation for the platform's visual design, with many describing their first impression as "grand," "beautiful," and "rich in color." This initial visual appeal quickly drew users in and engaged them in Browse the ethnic costumes (M=6.42, SD=1.14). They found the classification based on surface genes to be clear and easy to understand (M=6.13, SD=1.03), and noted that seeing the elements of pattern, color, material, and form increased their interest and curiosity (M=6.13, SD=0.90). This led them to select surface genes for in-depth exploration based on personal interest and characteristics. For example, P3 stated:

> *"I chose 'material' because I was curious about the material photos and wanted to know what they were exactly. For colors, I wanted to understand the distinction between cool, warm, and neutral, while for accessories and clothes, I chose based on my own gender."*

To further investigate the usefulness of translating surface-level genes, we asked participants about the difference between exploring through these costume elements versus the common method of exploring by ethnic group. Among the 24 participants, only one expressed a distinctly negative sentiment. The rest held positive or neutral views, with 10 participants explicitly pointing out that this classification method helped them focus more on the details and characteristics of the costumes, and many stated it could stimulate their desire to explore.

> *P17 noted: "I noticed some costume details and information I hadn't paid attention to before. I was impressed that fish skin could be used to make clothes, shoes, and bags, and the story of the Hezhen people using it for warmth broadened my knowledge."*
>
> *P14 found it to be "more detailed and helpful for users to focus on the differences and design features among the costumes."*

### 6.2.2 Rich Cultural Context as a Bridge to Deeper Understanding

After completing their free exploration of the costumes, participants indicated that the information on the detail pages was comprehensive and rich (M=6.08, SD=0.72). P5 expressed surprise at the details of the cultural information and craft techniques behind the costumes. These details guided the audience into the ethnic cultural atmosphere and helped them understand the culture from a richer dimension (M=6.04, SD=0.81).

> *P23 stated: "I have a very deep memory of the cultural background, residential composition, historical introduction, background stories, and the arts and entertainment aspects in the detailed description."*

### 6.2.3 AI Co-Creation as a Peak Experience with Room for Improvement

When asked about the best part of their experience, 15 participants identified the AI generation feature. They believed the generated content helped them better understand the cultural background of the costume (M=5.33, SD=1.63) and resonate with the inner-layer genes, such as the ethnic spirit.

> *"The quality of the generated content is high; the story has warmth and emotion," P24 commented.*

> *P23 also stated: "The generated story text met my expectations in terms of historical background, national sentiment, and the core values." This suggests that the prompts, designed based on the middle and inner-layer genes, successfully guided users to focus on the deep cultural values behind the costumes.*

However, participants also offered constructive criticism regarding the quality of the generated content, mainly focusing on image artifacts, a weak connection between the text and images, and a lack of depth or stylistic variety.

> *P7 noted "some image flaws and a weak correlation between the story and the image," while P15 mentioned that "the style of the generated content was too cartoonish" and expressed a desire for a feature to select different styles.*

Overall, participants showed great enthusiasm when discussing this feature. The experience design effectively sparked their exploratory interest and curiosity (M=5.46, SD=1.56), leading to a deeper understanding of the ethnic culture (M=5.33, SD=1.71).

### 6.2.4 From Experience to Identity: Fostering Affective Commitment and Behavioral Intent

**Affective Commitment** In the interviews, participants frequently expressed that they perceived the diversity of ethnic cultures. By gaining a deeper understanding of the cultural connotations of the costumes, they felt a greater sense of responsibility to protect and inherit this cultural heritage.

> *"Learning that our country has so many excellent cultures, although some are not well-known to the public, and that there are people dedicated to inheriting and promoting them, has enhanced my sense of identity with my national culture," P17 reflected.*

On a 1-7 scale, participants agreed that the cultural exploration and experience on the website deepened their identity with ethnic costumes (M=6.08, SD=1.25). They felt a deeper emotional connection after viewing the scenes generated from the stories (M=5.13, SD=1.48) and were better able to feel the ethnic emotion or spiritual value conveyed by the costumes (M=5.71, SD=1.08). The individual cultural identity and emotional affect measured by the questionnaire reflect a collective emotion that transcends the individual, arising from a connection to the culture and indicating the participants' sense of belonging and pride in their ethnic costume heritage.

**Behavioral Exploration**

Twenty-one participants expressed a willingness to pay more attention to and inherit ethnic costume culture after the experience. They indicated they were more willing to spend time learning about other ethnic cultures, such as their history, traditions, and customs (M=6.00, SD=1.14), and to recommend the website to friends and family (M=6.00, SD=1.22). From the interviews, we found that participants actively made behavioral commitments linked to their emotions.

In terms of exploratory intent, as P12 stated, "This experience introduced me to many ethnic costumes I had never heard of before, and it sparked an interest in learning more about these ethnic groups." Participants developed a strong curiosity about the history, customs, and cultural characteristics of the ethnic groups they explored. Some participants also expressed a willingness to learn about and inherit ethnic culture through practical actions.

> *"When these elements are reflected in related cultural and creative products, I will make an extra effort to learn about them. I think this is also an effective way to inherit Chinese culture," said P4.*

Regarding the willingness to share, most participants expressed a desire to share on social media, provided the quality of the generated content met their expectations (M=5.83, SD=1.46).

> *"This theme is very suitable for sharing on social media. It not only allows me to showcase my own creations but also enhances everyone's sense of identity and pride in our ethnic cultures," said P2.*

P6 believed this was a good way to understand and disseminate culture. More interestingly, participants even began to envision ways of sharing that went beyond the digital platform.

> *P4 shared an expectation for "physical postcards or similar creative products that could be printed for sharing."*

The participants' willingness to explore and share ensures that cultural identity is not limited to an expression of belief but helps to facilitate continuous and in-depth behavioral practice.

# 7 Discussion

## 7.1 Revisiting the Research Questions

Our research aimed to answer the core question: **How can we innovate exploration and experience methods based on a hierarchical cultural gene model to promote users'** **identification with ethnic costume culture?** Our findings collectively point to a clear answer: this goal can be achieved through a carefully designed, progressive experience path that guides users from cognitive exploration to affective co-creation. The effectiveness of this path is demonstrated across the following three interconnected stages:

**First, the "gene-first" exploration approach, which uses surface-layer genes as entry points, serves as a successful foundation for sparking users' initial motivation and curiosity**. Our research found that, unlike traditional viewing models centered on complete garments, presenting "genes" like patterns and colors as independent units guides users to focus on previously unnoticed details. P14 found this method to be "*more detailed and helpful for users to focus on the differences and design features among the costumes,*" while P3 initiated a personalized exploration based on their own curiosity and characteristics (e.g., curiosity about materials, selecting garments based on gender). The quantitative data supports this, showing that this approach significantly sparked users' exploratory interest (M=6.13, SD=0.90). This confirms that our "gene-first" strategy can effectively support Breadth Exploration, laying a solid foundation for subsequent in-depth engagement.

**Second, providing rich "cultural context" at the middle-layer gene level effectively transforms user curiosity into deep understanding.** Once attracted by surface-level details, the platform's narrative information on history, customs, and craft techniques becomes a critical bridge leading users from superficial viewing to genuine insight. Participant feedback indicates that this contextual information brought them "surprises" and "gains" in knowledge. For instance, P17 was impressed by "*the story of the Hezhen people using fish skin for warmth,*" stating that it "*broadened my knowledge.*" P23, likewise, recalled "*the cultural background, residential composition, and historical introduction*" in great detail. This indicates that an effective experience path must reconnect isolated cultural symbols with their rich socio-cultural backgrounds to support users' In-depth Exploration.

**Finally, the AI co-creation experience, which transforms the exploration of inner-layer genes into a personalized, affective practice, acts as the key catalyst for achieving cultural identity.** After completing the cognitive exploration phase, we invited users to participate in the co-creation of cultural narratives through the AI generation feature. Although evaluations of the generated content's quality were mixed, more than half of the participants considered it the "peak" of their experience. P24 felt the generated story had "*warmth and emotion,*" and P23 found it met expectations regarding "*historical background and national sentiment.*" This shift from explorer to co-creator is a core mechanism for promoting cultural identity. Our quantitative and qualitative data show that this experience translates directly into affective commitment. Users' sense of identity was significantly enhanced after the experience (M=6.08, SD=1.25), and qualitative interviews revealed their emotional leap from "knowing" to "pride" and "a sense of responsibility" (P17).

### 7.2   Implications for Theory and Design

The findings of this study offer new theoretical perspectives and practical pathways for the translation and revitalization of cultural heritage in the digital age. At a theoretical level, the core contribution of this research is its success in operationalizing abstract cultural theories into a designable and evaluable interactive experience framework. Our work extends the application of "Cultural Gene" theory from its previous use as a static analytical framework for physical products (Song & Zhan, 2020) to a generative principle capable of guiding the design of dynamic, personalized experiences. Simultaneously, our design practice provides a concrete digital exemplar for the identity formation theory of Crocetti et al. (2008). It clearly demonstrates how different interaction strategies can support users' Breadth Exploration (through the generous interface of surface genes) and In-depth Exploration (through the interpretation of middle-layer contexts and co-creation with inner-layer values), thereby providing a clear and feasible practical path for how to systematically promote the formation of cultural identity by guiding exploratory behaviors in a digital experience.

These theoretical insights directly inform actionable strategies for design practice. A core implication is the conceptual shift from object-centric replication to relationship-centric experience design. This means the focus of future digital heritage design should move beyond the high-fidelity reproduction of artifacts to meticulously crafting the user's exploratory and affective relationship with the cultural connotations. While the VisHanfu project effectively organizes knowledge exploration around an object's physical structure (Yu et al., 2024), our research proposes a more generative strategy: using the micro "cultural gene" as an entry point to guide users in autonomously discovering and constructing knowledge through a "from-point-to-surface" approach.

Establishing this new relationship often requires a core interactive metaphor, which we term a "cultural agent." The study by Mo et al. (2023), which ingeniously replaced modern "photo-taking" with the traditional craft of "rubbing" as the core interaction, is a prime example. Inspired by this, our research designs AI story generation as a narrative agent that connects the user to the inner-layer genes. The full potential of this narrative agent is ultimately realized through generative AI, positioning it as the user's "co-creative partner in cultural narrative." This not only supports the findings of Wang et al. (2023) that AI can empower users as creators but, more importantly, our research highlights the necessity of "scaffolded co-creation." The three-layer cultural gene model plays the critical role of a scaffold in this process, constraining and guiding the AI's output to ensure that its creativity does not deviate from the cultural origin and meaning, thus striking a delicate balance between innovation and inheritance.

### 7.3 Limitations and Future Work

This study has several limitations. First, the sample size of 24 participants is relatively limited, and they are mostly young people with similar educational backgrounds; the generalizability of the conclusions needs to be verified in a broader population. Second, as pointed out in the user feedback, the quality of the AI-generated content (e.g., image artifacts, stylistic uniformity) is a major bottleneck in the current experience, indicating significant room for optimization in our prompt engineering and the AI models we rely on.

Future work could expand in several directions:

1. Application and Extension of the Framework: Apply our "Three-Layer Cultural Gene" design framework to the digital innovation of other types of cultural heritage (e.g., architecture, artifacts, music) to test its universality.
2. Enrichment of Interaction Modalities: Drawing inspiration from the exploration of multi-sensory interaction in the Restoring Dunhuang Murals study (Sun et al., 2024), future work could introduce more immersive modalities such as VR/AR or haptic feedback to our platform to further enhance "affective connectedness."
3. Exploration of Social Co-creation Mechanisms: Inspired by the concept of large-screen, multi-user "digital rubbing" by Mo et al. (2023), future research could explore social features that support multiple users in collaboratively creating ethnic costume stories, allowing the construction of cultural identity to evolve from an individual experience to a community-shared one.
4. Deepening of AI Capabilities: Explore more advanced AI models and design more complex prompt strategies, for instance, allowing users to upload personal photos for style transfer or providing personalized story theme recommendations based on their Browse history.

# 8   Conlusion

As a vital carrier of ethnic identity, ethnic costume requires an innovative paradigm for its digital transmission that transcends static display. By constructing a hierarchical framework for cultural gene translation and integrating it with generative AI technology, this paper has designed and validated a digital platform that guides users through an in-depth cultural exploration and experience. The research results show that the platform can effectively enhance users' cultural cognition and affective identity. This study not only provides a systematic design strategy and practical exemplar for how to translate abstract cultural connotations into meaningful interactive experiences but also opens up new possibilities for the digital innovation of other forms of intangible cultural heritage, helping traditional culture to be revitalized in the new era.

*Interactive experience website URL: http://111.230.109.148:8080,*

*Login email: hcmx_sysu@163.com, Password: hcmxSYSU2025@*

## References

Bruner, J. (1992). Acts of meaning: Four lectures on mind and culture (Jerusalem-Harvard lectures). Harvard University Press.
Chu, Y. (2015). 中国服饰文化基因初探——为 APEC 会议领导人设计服装的思考 [A preliminary exploration of Chinese clothing's cultural genes—Thoughts on designing attire for APEC leaders]. 艺术设计研究 [Art and Design Research], (1), 30-34.
Crocetti, E. (2017). Self and identity formation as embedded in the social context. The Japanese Journal of Adolescent Psychology, 29(1), 1–16. https://doi.org/10.20688/jsyap.29.1_1

Crocetti, E., Rubini, M., & Meeus, W. (2008). Capturing the dynamics of identity formation in various ethnic groups: Development and validation of a three-dimensional model. Journal of Adolescence, 31(2), 207-222. https://doi.org/10.1016/j.adolescence.2007.09.002

Dawkins, R. (2016). The selfish gene. Oxford University Press.

Deng, L., Chen, B., Zhang, X., & Yang, X. (2018). 凉山彝族服饰文化基因提取及应用 [Extraction and application of cultural genes of Liangshan Yi ethnic costume]. 包装工程 [Packaging Engineering], 39(2), 270-275.

Devine, C. (2015). The museum digital experience: considering the visitor's journey. In MWA2015: Museums and the Web Asia.

Falk, J. H., & Dierking, L. D. (2013). Museum experience revisited. Left Coast Press.

Goodrum, A. A., & Martin, K. (1999). Bringing fashion out of the closet: Classification structure for the Drexel Historic Costume Collection. Bulletin of the American Society for Information Science and Technology, 25(4), 21-23. https://doi.org/10.1002/bult.133

Han, Z. (2010). 论国家认同、民族认同及文化认同——一种基于历史哲学的分析与思考 [On national identity, ethnic identity, and cultural identity: An analysis and reflection based on historical philosophy]. 北京师范大学学报(社会科学版) [Journal of Beijing Normal University (Social Sciences Edition)], (1), 106-113.

Hu, J. (2001). 中国少数民族的服饰文化 [The costume culture of China's ethnic minorities]. 广西民族研究 [Guangxi Ethnic Studies], (1), 62-68.

Jiang, Y., Xie, C., & Mao, J. (2023, October 9). The impact of identity construction and diversification of Chinese craftspeople on the design innovation of traditional handicrafts – a case study of Dong Brocade in Tongdao, Hunan. IASDR 2023: Life-Changing Design. https://doi.org/10.21606/iasdr.2023.420

Kalay, Y. E. (2007). Introduction: Preserving cultural heritage through digital media. In New heritage: New media and cultural heritage (pp. 17-26). Routledge.

Liu, B. (2007). 试析黔东南苗族服饰的文化特征 [A preliminary analysis of the cultural characteristics of Miao costumes in Southeast Guizhou]. 贵州民族研究 [Guizhou Ethnic Studies], 27(5), 61-66.

Luo, S., Wang, Y., Zhong, F., & Guo, Q. (2023). 创新设计转译文化基因的数字开发与传播策略研究 [Research on the digital development and communication strategy of translating cultural genes through innovative design]. 浙江大学学报(人文社会科学版) [Journal of Zhejiang University (Humanities and Social Sciences)], 53(1), 5-18.

Ma, S. (2019). "中华民族服饰文化研究" 的价值要义与方法构建 [The value essence and method construction of "Research on Chinese National Costume Culture"]. 艺术设计研究 [Art and Design Research], 15-18.

Marchionini, G. (2006). Exploratory search: from finding to understanding. Communications of the ACM, 49(4), 41-46.

Marcia, J. E. (1966). Development and validation of ego-identity status. Journal of Personality and Social Psychology, 3(5), 551–558. https://doi.org/10.1037/h0023281

Meeus, W. (2011). The Study of Adolescent Identity Formation 2000–2010: A Review of Longitudinal Research. Journal of Research on Adolescence, 21(1), 75–94. https://doi.org/10.1111/j.1532-7795.2010.00716.x

Meng, Z., & Liu, Y. (2015). 虚拟民族服饰博物馆的创建 [The creation of a virtual ethnic costume museum]. 北京服装学院学报（自然科学版） [Journal of Beijing Institute of Fashion Technology (Natural Science Edition)], 35, 9-14.

Mendoza, M. A. D., De La Hoz Franco, E., & Gómez, J. E. G. (2023). Technologies for the Preservation of Cultural Heritage—A Systematic Review of the Literature. Sustainability, 15(2), 1059. https://doi.org/10.3390/su15021059

Mo, Z., Li, D., & Ji, T. (2023, October 9). Inheriting the Intangible Cultural Heritage and embracing innovation:Digital Rubbing leads a new Experience of Audience Interaction in museums. IASDR 2023: Life-Changing Design. https://doi.org/10.21606/iasdr.2023.590

Pérez-Gandarillas, L., Manteca, C., & Yedra, Á. (2024). Conservation and protection treatments for cultural heritage: Insights and trends from a bibliometric analysis. Coatings, 14, 1027.

Phinney, J. S., & Ong, A. D. (2007). Conceptualization and measurement of ethnic identity: Current status and future directions. Journal of Counseling Psychology, 54(3), 271–281. https://doi.org/10.1037/0022-0167.54.3.271

Shao, L., & Shen, R. (2002). 中国美术大辞典 [The great dictionary of Chinese fine arts].

Song, X., & Zhan, B. (2020). 蒙古族服饰文化因子提取及设计应用 [Extraction and design application of Mongolian costume cultural factors]. 包装工程 [Packaging Engineering], 41(10), 325-330. doi:10.19554/j.cnki.1001-3563.2020.10.054.

Sun, T., Jin, T., Huang, Y., Li, M., Wang, Y., Jia, Z., & Fu, X. (2024). Restoring Dunhuang Murals: Crafting Cultural Heritage Preservation Knowledge into Immersive Virtual Reality Experience Design. International Journal of Human–Computer Interaction, 40(8), 2019–2040. https://doi.org/10.1080/10447318.2023.2232976

Thudt, A., Hinrichs, U., & Carpendale, S. (2012). The bohemian book-shelf: supporting serendipitous book discoveries through information visualization. In Proceedings of the SIGCHI conference on human factors in computing systems.

Umaña-Taylor, A. J., Quintana, S. M., Lee, R. M., Cross Jr., W. E., Rivas-Drake, D., Schwartz, S. J., Syed, M., Yip, T., Seaton, E., & Group, E. and R. I. in the 21st C. S. (2014). Ethnic and Racial Identity During Adolescence and Into Young Adulthood: An Integrated Conceptualization. Child Development, 85(1), 21–39. https://doi.org/10.1111/cdev.12196

Wang, D. (2003). 中华文明的文化基因与现代传承(专题讨论)中华文明的五次辉煌与文化基因中的五大核心理念 [The cultural genes of Chinese civilization and modern inheritance (symposium) The five glories of Chinese civilization and the five core concepts in cultural genes]. 河北学刊 [Hebei Academic Journal], (5), 130-134, 147.

Wang, P. (2018). 中国少数民族服饰文化研究 [Research on Chinese ethnic minority costume culture]. 内蒙古科学技术出版社 [Inner Mongolia Science and Technology Press].

Wang, S., & Bo, G. (2019). 清代服饰三蓝绣基因图谱研究 [A study on the gene map of Sanlan embroidery in Qing Dynasty costumes]. 丝绸 [Journal of Silk], (1).

Wang, S., Zhao, D., & Lu, S. (2023, October 9). Review: Design reshape the relationship between museum collections and visitors in digital age. IASDR 2023: Life-Changing Design. https://doi.org/10.21606/iasdr.2023.556

Wang, Z., Liu, F., Ran, C., & Zhang, M. (2023, October 9). AI Promotes the Inheritance and Dissemination of Chinese Boneless Painting——Research on Design Practice from Interdisciplinary Collaboration. IASDR 2023: Life-Changing Design. https://doi.org/10.21606/iasdr.2023.391

Wei, D., Zhou, X., Zhou, Y., & Wang, J. (2019). 基于多维特征的瑶族服饰数字化传播与潮流化研究 [Research on the digital dissemination and fashion trend of Yao nationality costumes based on multi-dimensional features]. 文化创新比较研究 [Comparative Study of Cultural Innovation], 3(17), 193-194.

Whitelaw, M. (2015). Generous interfaces for digital cultural collections. Digital Humanities Quarterly, 9(1).

Xu, D., Ren, S., & Yang, J. (2023). 基于扎根理论的绕家服饰纹样的系统性识别与文化蕴意研究 [Systematic identification and cultural connotation research of Raojia costume patterns based on grounded theory]. 丝绸 [Journal of Silk], 60(7), 116-123.

Yan, L., & Dong, B. (2023). 从"文化探索"到"文化体验":文化认同的形成机制及教育路径 [From "cultural exploration" to "cultural experience": The formation mechanism and educational path of cultural identity]. 全球教育展望 [Global Education], 52(1), 32-46.

Yu, J., & Zhu, W. (2024). 大数据驱动的生成式 AI 在服装设计中的应用——以 Midjourney 为例 [Application of big data-driven generative AI in fashion design—A case study of Midjourney]. 丝绸 [Journal of Silk], 61(9), 20-27.

Yu, M., Zeng, L., Du, X., Sheng, J., Liao, Q., & Liu, Y.-J. (2024). VisHanfu: An Interactive System for the Promotion of Hanfu Knowledge via Cross-Shaped Flat Structure. Proceedings of the 32nd ACM International Conference on Multimedia, 3047–3055. https://doi.org/10.1145/3664647.3681353

Zhao, C. (2008). 论文化基因及其社会功能 [On cultural genes and their social functions]. 河南社会科学 [Henan Social Sciences], 16(2), 50-52.

Zhao, H. (2021). 文化基因研究缘起、进展与未来研究思考综述 [A review of the origin, progress, and future research of cultural gene studies]. 中国传媒大学学报(自然科学版) [Journal of Communication University of China (Natural Science Edition)], 28(5), 1-10.

Zhao, J., & Wang, G. (2011). 关于数字化民族服饰图案文化的保护与研究 [On the protection and research of digital ethnic costume pattern culture]. 美术教育研究 [Art Education Research], (12), 20-21.

**About the Authors:**

**Ma Xiaofan:** She is a master candidate in the School of Journalism and Communication at Sun Yat-sen University. Her research lies at the intersection of HCI, Human–AI Collaboration, and Cultural Experience Design. She designs and studies AI-mediated interactions that enhance human understanding, creativity, and communication in interdisciplinary and cultural contexts.

**Yan Lirong:** She is a master's student in Interaction Design at Sun Yat-sen University. With a research focus on Human-Computer Interaction and Interaction Design in cultural contexts, she is dedicated to exploring how interactive technologies can enhance cultural heritage preservation and user engagement in digital cultural experiences.

**Weijia Zhao:** She is currently a master candidate at Sun Yat-sen University. She is committed to exploring innovative design approaches for Intangible Cultural Heritage (ICH) preservation and the integration of explainable design principles in creative practices.

**Zeng Weiping:** He is currently a master candidate at the School of Journalism and Communication, Sun Yat-sen University. With a background and skills in website design and development, he is dedicated to exploring website design and development with excellent user experiences.

**Wu Huiyue:** Huiyue Wu is currently a Full Professor at Sun Yat-sen University, Guangzhou, China, where he is also the vice dean of the School of Journalism and Communication. He received the Ph.D. degree in computer science from Institute of Software, the Chinese Academy of Sciences, China, in 2010. His research interests include human-computer interaction, virtual and augmented reality. He is the author of seven books and more than 60 publications in the field of Human-Computer Interaction (e.g., IJHCS, TVCG).

**Acknowledgement:** This study builds upon a group project from the required course 'Research on User Behavior and Human-Computer Interaction'. The initial work was later independently expanded and refined by the first author, Xiaofan Ma. The authors would like to thank the instructors and peers for their guidance and feedback during the course, as well as the institution and participants who contributed to the user study.